\documentclass[twocolumn]{aastex631}

\begin{document}

\title{Blowing star formation away in AGN Hosts (BAH) - II. Investigating the origin of the H$_2$ emission excess in nearby galaxies with  JWST MIRI}

\author[0000-0003-0483-3723]{Rogemar A. Riffel}
\author[0009-0005-0583-5773]{Gabriel L. Souza-Oliveira}
\author[0000-0003-3667-9716]{José Henrique Costa-Souza}
\affiliation{Departamento de F\'isica, CCNE, Universidade Federal de Santa Maria, Av. Roraima 1000, 97105-900,  Santa Maria, RS, Brazil}

\author[0000-0001-6100-6869]{Nadia L. Zakamska}
\affiliation{Department of Physics \& Astronomy, Johns Hopkins University, Bloomberg Center, 3400 N. Charles St, Baltimore, MD 21218, USA}

\author[0000-0003-1772-0023]{Thaisa Storchi-Bergmann}
\author[0000-0002-1321-1320]{Rog\'erio Riffel}
\affiliation{Departamento de Astronomia, IF, Universidade Federal do Rio Grande do Sul, CP 15051, 91501-970, Porto Alegre, RS, Brazil}

\author[0000-0002-6570-9446]{Marina Bianchin}
\affiliation{Department of Physics and Astronomy, 4129 Frederick Reines Hall, University of California, Irvine, CA 92697, USA}



\begin{abstract}
We use James Webb Space Telescope (JWST) Mid-Infrared Instrument (MIRI) medium-resolution spectrometer (MRS) observations of 3C 293 (UGC\:8782), CGCG\:012-070 and NGC\:3884 to investigate the origin of the H$_2$ emission. These three nearby Active Galactic Nucleus (AGN) hosts are known to present H$_2$ emission excess relative to star-forming galaxies, as traced by the H$_2$ S(3)/PAH$_{\rm 11.3\mu m}$ line ratio. We define the kinematically disturbed region (KDR) by the AGN and the virially dominated region (VDR) based on the H$_2$ line widths, using the $W{\rm 80}$ parameter. From the correlations between $W{\rm 80}$ and H$_2$ S(3)/PAH${\rm 11.3\mu m}$, as well as the higher H$2$ S(5)/H$2$ S(3) and [Fe\:{\sc ii}]${\rm 5.34\: \mu m}$/PAH${\rm 11.3\mu m}$ ratios and flatter power-law temperature distributions observed in the KDR, we conclude that the H$_2$ emission in the KDR is primarily driven by shock-heated gas. For 3C\:293, the KDR is co-spatial with the radio core, indicating that the origin of the shocks is the interaction of the radio jet with the interstellar medium, which is also responsible for the observed molecular and ionized gas outflows in this source. The other galaxies are weak radio sources; however, due to the lack of high-resolution radio images, we cannot rule out low-power jets as the origin of the shock-heated H$_2$. Our results indicate that the excess H$_2$ emission  excess is associated to shock heating of the gas, generated by outflows or by the interaction of the radio jet with the ambient gas. 

\end{abstract}


\section{Introduction}

The molecular hydrogen (H$_2$) is the most abundant molecule in the Universe and the raw fuel of star formation \citep[e.g.][]{Omont07}. However, the origin of H$_2$ emission and dynamics in extreme environments, such as active galactic nuclei (AGN), remains poorly understood. Atacama Large Millimeter Array (ALMA) observations have mapped the cold molecular gas ($T\lesssim$100 K) in nearby AGN hosts, uncovering signs of inflows and outflows \citep[e.g.][]{almudena19}. These studies use tracer molecules such as CO and HCN, without directly mapping H$_2$ emission or extending beyond the cold molecular phase. Ground-based near-infrared integral field unit (IFU) observations provide direct insights into the emission of the H$_2$ molecule (using the vibrational H$_2$ transitions, mostly in the K-band), but are confined to the hot molecular gas  ($T\gtrsim$1000 K), which represents only a hot skin of the molecular gas content in the central region of nearby galaxies \citep{rogemar21_survey,rogemar23_extended_kin}. The physics and dynamics of warm molecular gas phase (100 K $\lesssim T\lesssim$ 1000 K) is now accessible through observations with the James Webb Space Telescope (JWST), which have the spatial and spectral resolutions necessary to resolve the emission structure and kinematics of ro-vibrational lines of H$_2$ in the mid-infrared (MIR).

Indirect evidence of the AGN's effect on the origin of H$_2$ MIR is provided through observations with the Spitzer telescope. \citet{lambrides19} conducted a compreheensive analysis on the spectra of a sample comprising 2015 galaxies observed with Spitzer's Infrared Spectrograph, encompassing both AGN hosts and star-forming galaxies. They derived the excitation temperatures of the H$_2$ and found that AGN-dominated galaxies exhibit temperatures, on average, 200 K higher than those without AGN, consistent with previous results in smaller sub-samples \citep{Petric18}. In addition, \citet{lambrides19} found that AGN hosts present an H$_2$ emission excess over what is expected for pure star formation, parameterized by the H$_2$S(3)\:9.665$\mu$m/PAH\:11.3$\mu$m. \citet{rogemar20_spitzer} conducted a follow-up study investigating the properties of optical emission lines by cross-correlating \citet{lambrides19} sample with the third-generation Sloan Digital Sky Survey (SDSS-III) database \citep{gunn06,eisenstein11,smee13,Thomas13}. This comparison reveals a strong association between the H$_2$ emission excess and shock tracers in neutral gas, such as high values of [O\:{\sc i}]$\lambda$6300 velocity dispersion and high [O\:{\sc i}]$\lambda$6300/H$\alpha$ line ratio \citep[e.g.][]{allen08,ho14,dors21_suma}. This indicates that galaxies with a high H$_2$S(3)\:9.665$\mu$m/PAH\:11.3$\mu$m ratio may experience an excess of H$_2$ emission due to shocks resulting from the interaction of outflows with the ambient gas. This is further reinforced by simulations, which suggest that molecules could be generated within the outflow, exhibiting excitation properties indicative of shock heating \citep{richings18a,richings18b}.

The use of JWST, in combination with the Mid-Infrared Instrument (MIRI) medium-resolution spectrometer (MRS), enables detailed studies of the emission and dynamics of warm molecular gas in nearby galaxies, utilizing the MIR rotational H$_2$ lines. Results obtained with MIRI in the study of molecular gas in nearby galaxies include:  increased emissions of warm and hot H$_2$ as well as ionized gas, which coincide with the intersection of the jet with dust lanes in the disk of NGC 7319 \citep{Pereira-Santaella2022}; regions of increased H$_2$ velocity dispersion between the nucleus and the star-forming ring in NGC\:7469, attributed to shocks generated by ionized outflows \citep{U22}; a wealth of interacting structures among various gas phases in the intergalactic medium in Stephan's Quintet, with strong MIR H$_2$ line emissions from warm gas generated by the fragmentation of dense cold molecular clouds, followed by mixing in the post-shock gas \citep{Appleton23}; a collimated outflow of warm molecular gas originating from the southern nucleus of the ultra-luminous infrared galaxy NGC\:3256, which is co-spatial with the outflow in cold gas observed by ALMA \citep{Sakamoto14}. The warm to cold mass fraction is roughly 4\%; a large fraction of warm molecular gas, accounting for 75\% of the total molecular gas mass, in M83 \citep{Hernandez2023}; powerful jet-driven outflows in warm molecular gas in  the inner few kpc of nearby ($z<0.1$) radio-loud  AGN \citep{Leftley24,henrique24};
evidence of shock excitation of H$_2$ and detection of warm molecular outflows within the ionization cones of the nearby Seyfert galaxy NGC\:5728 \citep{Davies24}. \citet{ogle25} used MIRI, along with near-infrared and optical integral field spectroscopy, to study Cygnus A, revealing a density-stratified NLR shaped by the interaction of the radio jet with the interstellar medium, and multi-phase biconical outflows reaching velocities of 600--2000 km\:s$^{-1}$.

Here, we use JWST MIRI MRS to investigate the origin of the H$_2$ emission excess in three nearby active galaxies: 3C\:293, CGCG\:012-070 and NGC\:3884. These targets were selected from \citet{rogemar20_spitzer},  identified as the most extreme objects in terms of H$_2$S(3)\:9.665$\mu$m/PAH\:11.3$\mu$m, [O\:{\sc i}]$\lambda$6300 velocity dispersion and high [O\:{\sc i}]$\lambda$6300/H$\alpha$ line ratio, indicating they may be potential hosts of molecular gas outflows and shocks. This paper is structured as follows: Section 2 introduces the targets and the data, Sec. 3 presents our results, which are discussed in Sec. 4. Finally, our main conclusions are summarized in Sec. 5.

\section{Targets and data}
\subsection{The galaxies}

{\bf 3C\,293 (UGC\,8782)}  presents a complex optical morphology, leading to its classification 
 both as a spiral galaxy \citep[e.g.][]{sandage66,rc1991} and as a dusty elliptical \citep[e.g.][]{Ebneter1985,Tremblay2007}. It is located at a redshift $z = 0.045$ \citep{sandage66}, is host of a Low-Ionization Nuclear Emission Region \citep[LINER;][]{veron06} and  is a radio loud source classified as Fanaroff-Riley (FR) II \citep{Fanaroff1974,liu02}, displaying a double-double radio structure. The inner radio lobes are observed at scales of $\sim$4 kpc, aligned along the east-west direction, while the outer lobes span approximately 200 kpc along the northwest-southeast direction \citep[e.g.,][]{machalski16}. Sub-arcsecond radio imaging of the 3C\:293 indicates it has gone through multiple activity epochs, with a compact young nuclear steep spectrum (CSS) source, and it is classified as a jet with short interruption time periods \citep{Kukreti22}.
 
In the optical, 3C\:293 presents dust filaments spanning scales from 10$^2$--10$^3$ pc, which have been interpreted as a result of a previous merger event \citep[e.g.][]{martel99}.   Outflows co-spatial with the inner radio jet have been observed in ionized gas  \citep{emonts05,mahony16,rogemar23_ugc} with velocities of $\sim$1000 km\,s$^{-1}$,  and in neutral hydrogen \citep{morganti03} with velocities of up to 1400 km\,s$^{-1}$. The ionized gas emission in the inner 2 kpc of this galaxy is primarily attributed to photoionization by the central AGN, with an additional contribution from shocks for the outflow component, due to the interaction of the radio jet with the surrounding gas \citep{rogemar23_ugc}.  Previous Spitzer observations of 3C 293 reveal strong jet-shocked H$_2$ emission \citep{Ogle10}, indicating that the radio jet drives both the outflow and the heating of the warm H$_2$ gas \citep{Guillard12}.  While the outflows are not detected in cold molecular gas \citep{labiano14}, \citet{henrique24} report the detection of outflows in warm H$_2$, using JWST MIRI MRS data. The warm molecular outflows are observed co-spatial with the radio core and these authors estimated a mass-outflow rate of $\approx$6\:M$_\odot$\:yr$^{-1}$, which is one order of magnitude larger than the ionized gas outflows rate. If sustained, these outflows are capable of expelling all warm molecular gas from the central region of the galaxy in just 1 Myr \citep{henrique24}.

{\bf CGCG\:012-070} is a highly inclined Sb spiral galaxy, with axis ratio of $b/a\approx0.3$ \citep{Kautsch2006}, located at a redshift $z = 0.048$ \citep[e.g.][]{Nair2010}. Its nuclear nuclear activity is classified as a Seyfert 2 AGN  \citep[e.g.][]{veron06}. This galaxy presents faint radio emission, with a 1.4 GHz luminosity of $L_{\rm 1.4}\approx2.4\times10^{22}$ W\:Hz$^{-1}$ measured from the Faint Images of the Radio Sky at Twenty centimeters (FIRST) survey \citep{Becker1995,Lofthouse18,Lin18}. The nuclear optical spectrum CGCG\:012-070 exhibits intense emission lines, which can be represented by two kinematic components, a broad base and a narrow component, suggesting the presence of an outflow in ionized gas \citep{gunn06}. This is further confirmed by IFU observations, which reveal that the broad component is associated with a nuclear outflow, while the narrow component refers to the emission from the gas in the disk (Ramos Vieira, priv. comm.).

{\bf NGC\,3884} (UGC\:6746) is the closest of the three galaxies, with a redshift of $z = 0.023$ \citep[e.g.]{Rines2016}, classified as a Sab spiral \citep{Tully2000}. Its nuclear activity has been classified as a LINER with a faint H$\alpha$ emission from the broad line region \citep{veron06,rogemar_n3884}.
The ionized and neutral gas kinematics in the inner $\sim1.4$~kpc of NGC\:3884 present three kinematic components: a disc component similar to that observed for the stars; an inflowing gas component or gas externally acquired in previous merger events, observed as a twist in the kinematic position angle of the gas velocity field in the inner $\sim340$ pc; an outflow detected as broad components to the emission lines ($\sigma\sim$250--400\,km\,s$^{-1}$), with a maximum mass outflow rate of 0.25$\pm$0.15 M$_\odot$\,yr$^{-1}$ \citep{cazzoli18,hermosa-munoz22,rogemar_n3884}. This galaxy presents a faint radio source with $L_{\rm FIRST}\approx1.01\times10^{22}$ W\:Hz$^{-1}$ \citep{Lofthouse18,Lin18}.

\begin{figure*}
    \centering
    \includegraphics[width=0.79\textwidth]{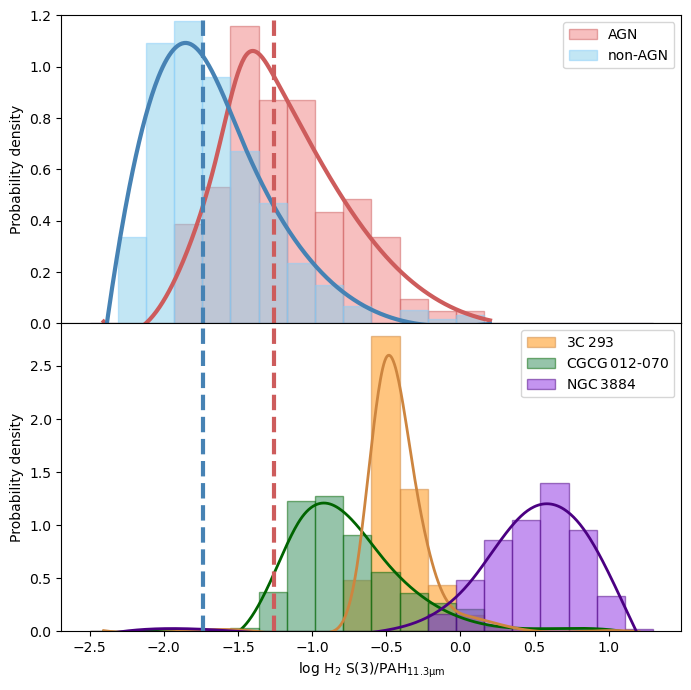}
    \caption{The top panel shows distributions of the H$_2$ S(3)/PAH$_{\rm 11.3\mu m}$ flux line ratios for AGN (light red) and non-AGN (light blue) hosts, based on the Spitzer data integrated over the entire galaxies, compiled by \citet{lambrides19}. The bottom panel shows the distribution of the observed H$_2$ S(3)/PAH$_{\rm 11.3\mu m}$ flux line ratios per spaxel for CGCG\:012-070 (green), 3C\:293 (orange) and NGC\:3884 (purple). The y-axis represents the probability density of values for each distribution, so that the integral of the area under the histogram equals 1. The solid lines show the smoothed distributions for each histogram, while the dashed vertical lines show the median values of H$_2$ S(3)/PAH$_{\rm 11.3\mu m}$  observed for AGN (red) and non-AGN (blue) hosts using the Spitzer dataset.  }
    \label{fig:spitzer}
\end{figure*}

\subsection{The data and measurements}

We use MIR JWST spectroscopic data obtained with the Mid-Infrared Instrument medium-resolution spectrometer \citep[MIRI MRS;][]{2015PASP..127..646W,Labiano21,2023A&A...675A.111A}, as part of a Cycle 1 JWST proposal (Prop. ID: 1928, PI: Riffel, R. A.).  We used a dither pattern consisting of four points, incorporating 30 groups with a single integration per dither position. Utilizing the slow readout mode, our total exposure time on the each source amounted to 1 hour and 35 minutes, plus 24 minutes of background exposure.  For 3C\:293 and CGCG\:012-070, we used the sub-bands Short and Long, while for NGC\,3884, the sub-bands Medium and Long were used.  These sub-bands were selected to include at least the H$_2$ S(1)\,17.03\,$\mu$m and H$_2$ S(3)\,9.665\,$\mu$m emission lines in all galaxies, used to trace the warm molecular gas kinematics and essential to probe the gas temperature and masses.  The observed spectral range also includes PAH features (at 11.3 and 6.2 $\mu$m), that can be used to map the H$_2$/PAH ratio, as well as emission lines from the ionized gas.  

Here, we focus mainly in the 9.5--12\:$\mu$m spectral region, which falls in the channel 2 Long sub-band, and use the H$_2$ S(3)/PAH$_{\rm 11.3\mu m}$ emission-line ratio to investigate the H$_2$ emission excess observed in these galaxies. The angular resolution of the datacubes is approximately 0.32 arcsec as given by the full width at half maximum (FWHM) of the JWST Point Spread Function (PSF), as measured from datacube for the star HD 2811.  We incorporate to our analysis other H$_2$ rotational lines and the [Fe\,{\sc ii}]$_{\rm 5.34\: \mu m}$ which contribute to our understanding of the the H$_2$ excitation source.
The data processing was conducted using the JWST Science Calibration Pipeline \citep{bushouse_2024}, version 1.13.4, along with the reference file \texttt{jwst\_1188.pmap}. Additionally, optimized routines were employed for outlier detection and removal of artifacts.  A detailed description of the observations and data reduction procedure is presented in \citet{gabriel24}.

We employed the {\sc ifscube} Python package \citep{Ruschel-Dutra21} to fit the profiles of the strongest emission lines by Gauss-Hermite functions, in a continuum subtracted cube. The continuum subtraction was made by applying a double locally weighted scatterplot smoothing (lowess) for each spectral interval in the observed spectrum, using 30\% and 10\% of the data points in this order, followed by a linear interpolation. Although lowess smoothing does not properly constitute a fitting method, it has shown to be effective and a simpler solution than polynomial fitting, given the complexity of the spectra and the resampling noise of spectral cubes.  To initialize the emission line fitting process, we start by fitting the the nuclear spaxel. After successfully fitting the emission line profiles for the nucleus, these parameters serve as initial guesses for neighboring spaxels, following a spiral loop pattern. In the {\sc ifscube} code, we set the {\it refit} parameter to utilize the best-fit parameters from spaxels located within distances less than 0.3\:arcsec as the starting point for subsequent fits. 

The IFSCUBE code generates a data cube that contains the best-fit parameters. These parameters are then used to map the distributions of emission line flux and their kinematics.  In addition, we use the {\sc ifscube} code to construct maps of the $W_{\rm 80}$ parameter, corresponding to the width of the line profile containing 80 per cent of its total flux. Finally, PAH$_{\rm 11.3\mu m}$ flux maps are obtained by direct integration of the observed emission feature, after the subtraction of the underlying continuum. In all maps, we exclude regions where the emission line amplitude is lower than 3 times the standard deviation of the continuum in a nearby spectral range, free of emission lines.

\section{Results}

\begin{figure*}
    \centering
    \includegraphics[width=0.75\textwidth]{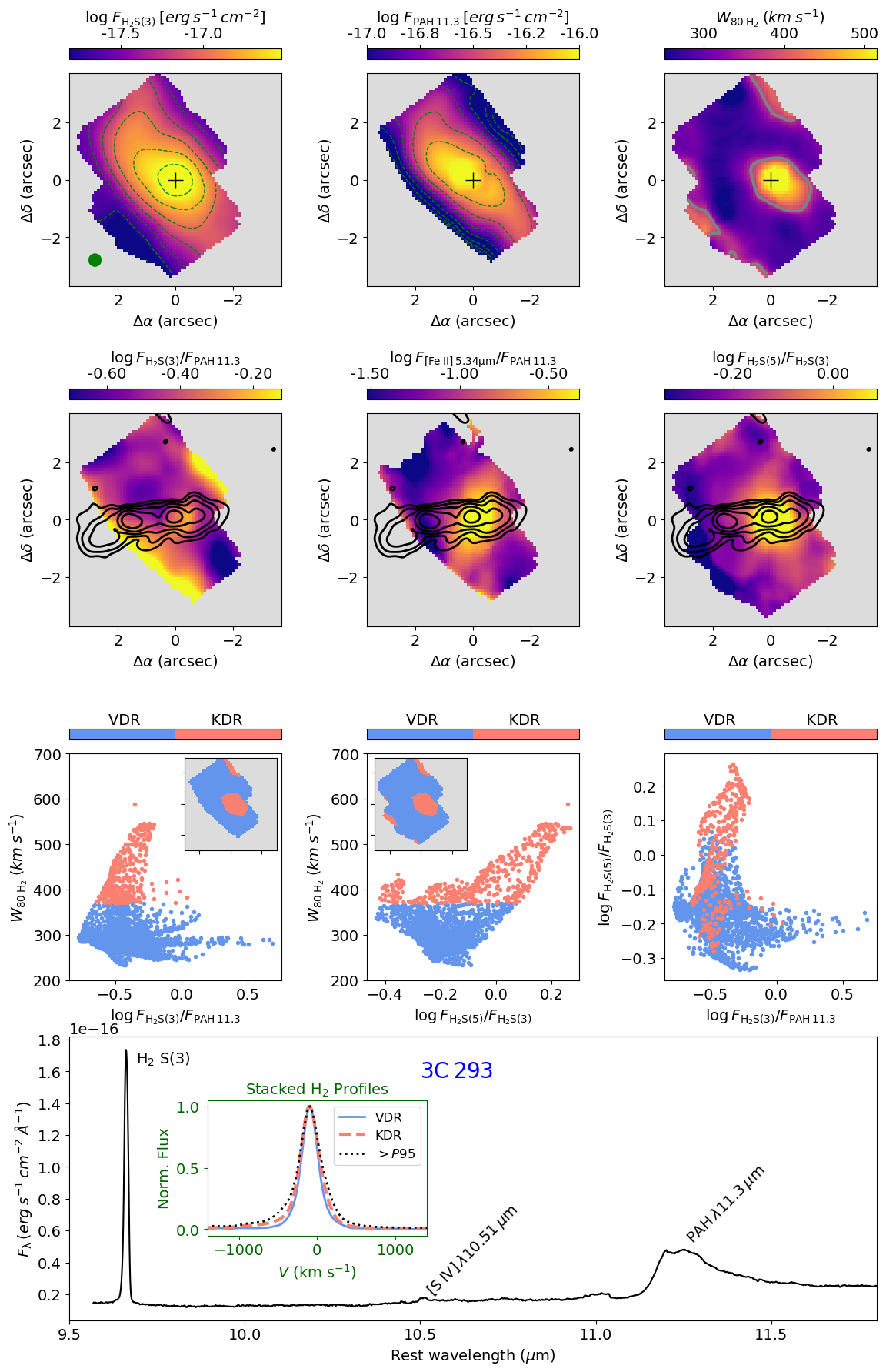}
    \caption{Results for 3C\:293. The top panels show, from left to right, the H$_2$ S(3) and PAH$_{\rm 11.3\mu m}$ flux maps,  and the $W_{\rm 80}$ map for the H$_2$ S(3) emission line. The green filled circle shows the PSF FWHM and the dashed gray contours show flux levels from each image. The contours in the $W_{\rm 80}$ map show the threshold used separate the virially dominated (VDR) and kinematically disturbed regions (KDR). The central plus signs represent the position of the peak of the continuum emission. The 2nd row shows the H$_2$ S(3)/PAH$_{\rm 11.3\mu m}$, [Fe\,{\sc ii}]$_{\rm 5.34\: \mu m}$/PAH$_{\rm 11.3\mu m}$ and H$_2$ S(5)/H$_2$ S(3) ratio maps. The contours  are from the 1360 MHz radio image of 3C\:293 from \citet{Kukreti22}.  The 3rd row show plots of $W_{\rm 80 H_2}$ vs. log\:H$_2$ S(3)/PAH$_{\rm 11.3\mu m}$, $W_{\rm 80 H_2}$ vs. log\:H$_2$ S(5)/H$_2$ S(3) and log\:H$_2$ S(5)/H$_2$ S(3) vs.  log\:H$_2$ S(3)/PAH$_{\rm 11.3\mu m}$. The insets indicate the KDR and VDR.   The bottom panel shows a spectrum integrated within an aperture of 2.5 arcsec radius, covering the spectral region from 9.5 to 11.8~$\mu$m. The H$_2$ S(3), [S\:{\sc iv}]$\lambda10.51\:\mu$m and PAH$_{\rm 11.3\mu m}$ features are labeled. The inset shows stacked profiles for the H$_2$ S(3) emission line in the  VDR and KDR.  In all maps, the gray regions correspond to locations where the corresponding emission line is not detected above $3\:\sigma$ of the continuum noise level or is not covered by the observed field of view. }
    \label{fig:ugc}
\end{figure*}

Figure~\ref{fig:spitzer} presents the log\:H$_2$ S(3)/PAH$_{\rm 11.3\mu m}$ emission-line ratio distributions for AGN (in red) and non-AGN (in blue) hosts as obtained from the Spitzer data, using the fluxes from \citet{lambrides19}. In this plot, AGN are defined as those objects with $6.2\:\mu$m PAH equivalent width, EQW$_{\rm 6.2\:\mu m} < 0.27\:\mu$m, while non-AGN have  EQW$_{\rm 6.2\:\mu m} > 0.27\:\mu$m \citep[e.g.][]{Stierwalt13}. The total number of AGN in the sample is 109, with a median value of log\:H$_2$ S(3)/PAH$_{\rm 11.3\mu m}=-1.26$, while the median value for the 313 non-AGN sample is log\:H$_2$ S(3)/PAH$_{\rm 11.3\mu m}=-1.74$. These values are indicated by the red and blue vertical dashed lines for AGN and non-AGN hosts, respectively. The bottom panels of Fig.~\ref{fig:spitzer} show the log\:H$_2$ S(3)/PAH$_{\rm 11.3\mu m}$ distributions for 3C\:293, CGCG\:012-070 and NGC\:3884, obtained with JWST. As can be observed, the three galaxies present higher values than the median one observed in the Spitzer AGN sample, with the highest ratios seen for NGC\:3884.

In Figure~\ref{fig:ugc}, we present the results for 3C\:293. The bottom panel shows an integrated spectrum in the 9.5--12\:$\mu$m spectral region, within a circular aperture of 2.5 arcsec radius, centered at the galaxy nucleus, defined as the position corresponding to the peak of the continuum emission. This galaxy presents strong H$_2$ S(3) and PAH$_{\rm 11.3\mu m}$ features, with extended emission observed over the whole MIRI MRS field of view (FoV), as seen in the first two panels of the top row. The H$_2$ emission peak is observed at the galaxy nucleus and is more elongated along the northeast-southwest direction (orientation of the galaxy's major axis), revealing spiral arms to both sides of the nucleus. The PAH emission is also extended along the same direction, showing clearly the spiral arms. It can be noticed that the peak of the PAH emission is slightly displaced to the east of the position where H$_2$ emission peaks, which is co-spatial with the MIR emission center. The PAH emission peak is coincident with the location of the optical center of 3C\:293. As already noticed in \citet{rogemar23_ugc} and \citet{henrique24}, the optical center of emission displaced about 0.6 arcsec from the MIR emission center, likely due to higher extinction in the optical continuum. 

The top-right panel of Fig.~\ref{fig:ugc} shows the $W_{\rm 80}$ map for the H$_2$ S(3) emission line  ($W_{\rm 80\:H_2}$) for 3C\:293. The highest values of up to $\sim$500\:km\:s$^{-1}$ are observed surrounding the galaxy's nucleus, slightly extending to the west of it. Some high $W_{\rm 80\:H_2}$ values are also seen in regions close to the border of the MIRI MRS FoV, mainly to the northwest and east/southeast of the nucleus. The lowest values of $W_{\rm 80\:H_2}\approx200$\:km\:s$^{-1}$ are observed surrounding the nucleus at distances of $\sim$1.5 arcsec from it and in regions along the spiral arms seen in the H$_2$ and PAH flux distributions. The $W_{\rm 80\:H_2}$ values presented here are not corrected by instrumental broadening \citep[$\sigma_{\rm ins}\approx$40 km\,s$^{-1}$;][]{Argyriou23}, which is negligible for the range of values observed. 

The left panel in the second row of Fig.~\ref{fig:ugc} shows the H$_2$ S(3)/PAH$_{\rm 11.3\mu m}$ line ratio map for 3C\:293. The highest values of up to log\:H$_2$/PAH$\approx0$  are observed at the nucleus and in regions farther away from it along the southeast-northwest direction. The lowest values of log\:H$_2$ S(3)/PAH$\approx-0.7$ are seen mainly along the northeast-southwest direction, along the galaxy's disk major axis. For comparison, the middle panel of the second row shows the [Fe\,{\sc ii}]$_{\rm 5.34\: \mu m}$/PAH$_{\rm 11.3\mu m}$ line ratio map, which displays similar structures as those seen in the H$_2$\;S(3)/PAH$_{\rm 11.3\mu m}$ ratio map.  The [Fe\,{\sc ii}] emission is usually associated with shocks in partially ionized gas, that destroy dust grains releasing the iron into the gaseous phase \citep[e.g.][]{sb2009_n4151,rogerio13,rogemar_n1068,hill14}. The right panel of this row shows the log\:H$_2$ S(5)/H$_2$ S(3) ratio map, a tracer of the gas temperature. The highest ratios (highest temperatures) are observed in regions co-spatial with locations where the highest H$_2$\:S(3)/PAH$_{\rm 11.3\mu m}$ values are seen. It is evident that the highest values of the three ratios are observed in regions co-spatial with the radio core, as indicated by the black contours representing the 1360 MHz radio image of 3C\:293 from \citet{Kukreti22}.

The third row of  Fig.~\ref{fig:ugc} show plots of $W_{\rm 80\:H_2}$ vs log\:H$_2$ S(3)/PAH$_{\rm 11.3\mu m}$, $W_{\rm 80\:H_2}$ vs. log\:H$_2$ S(5)/H$_2$ S(3) and log\:H$_2$ S(5)/H$_2$ S(3) vs. log\:H$_2$ S(3)/PAH$_{\rm 11.3\mu m}$, which will be further used to discuss the origin of the H$_2$ emission in the next section. 

\begin{figure*}
    \centering
    \includegraphics[width=0.75\textwidth]{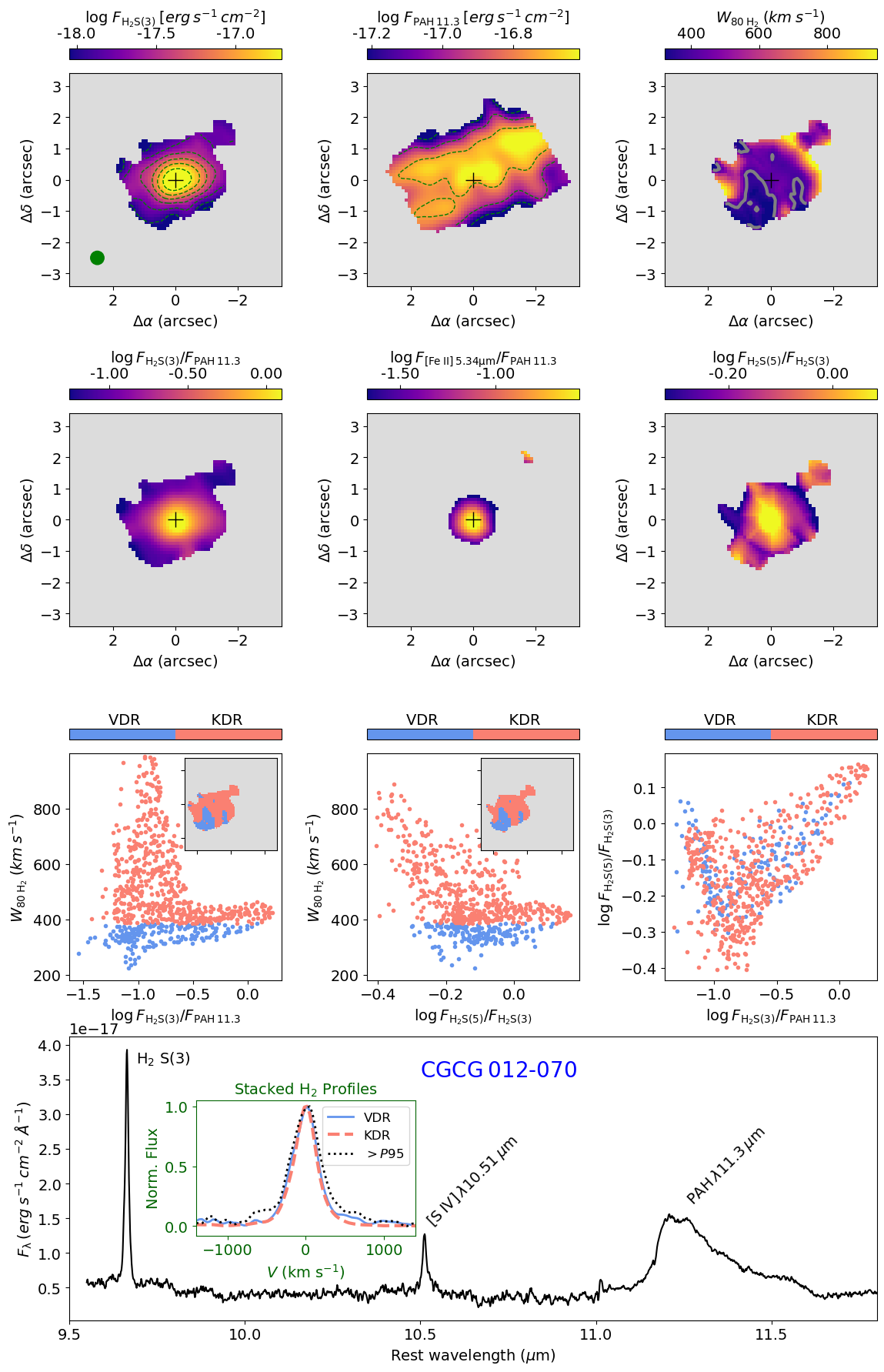}
    \caption{Same as Fig.~\ref{fig:ugc}, but for CGCG\:012-070. Radio contours are not shown as there are no high resolution radio data available for this galaxy.}
    \label{fig:cgcg}
\end{figure*}

Figure~\ref{fig:cgcg} shows the results for CGCG\:012-070. This galaxy displays strong H$_2$ and PAH emission, as revealed by the integrated spectrum shown in the bottom panel of Fig.~\ref{fig:cgcg}. The relative intensity of the [S\:{\sc iv}]10.51\:$\mu$m emission line is larger, as compared to 3C\:293. The H$_2$ emission is extended over the inner $\sim$1.8 arcsec, peaking at the nucleus and slightly more elongated to the northwest. The PAH emission is elongated along the southwest-northwest direction following the orientation of the galaxy's major axis. The highest PAH flux levels are observed at the north side of the nucleus and a ``V-shaped'' structure is observed close to the edge of the FoV to the southeast. The highest  H$_2$ S(3)/PAH$_{\rm 11.3\mu m}$ values, of $\sim$1.25, are observed at the nucleus, decreasing with the distance from  it. A similar behaviour is observed for the [Fe\,{\sc ii}]$_{\rm 5.34\: \mu m}$/PAH$_{\rm 11.3\mu m}$ line ratio map, with the largest values reaching [Fe\,{\sc ii}]$_{\rm 5.34\: \mu m}$/PAH$_{\rm 11.3\mu m}\approx1.0$. The highest values of the H$_2$ S(5)/H$_2$ S(3) flux ratio are concentrated in the innermost region and gradually decrease with increasing distance from the nucleus.  The  $W_{\rm 80\:H_2}$ map for this galaxy presents values of up to $\sim$900~km\:s$^{-1}$, mostly in regions away from the nucleus, but a linear structure of intermediate values ($\sim$600~km\:s$^{-1}$) is also observed along the northeast-southwest direction, crossing the galaxy's nucleus.

\begin{figure*}
    \centering
    \includegraphics[width=0.75\textwidth]{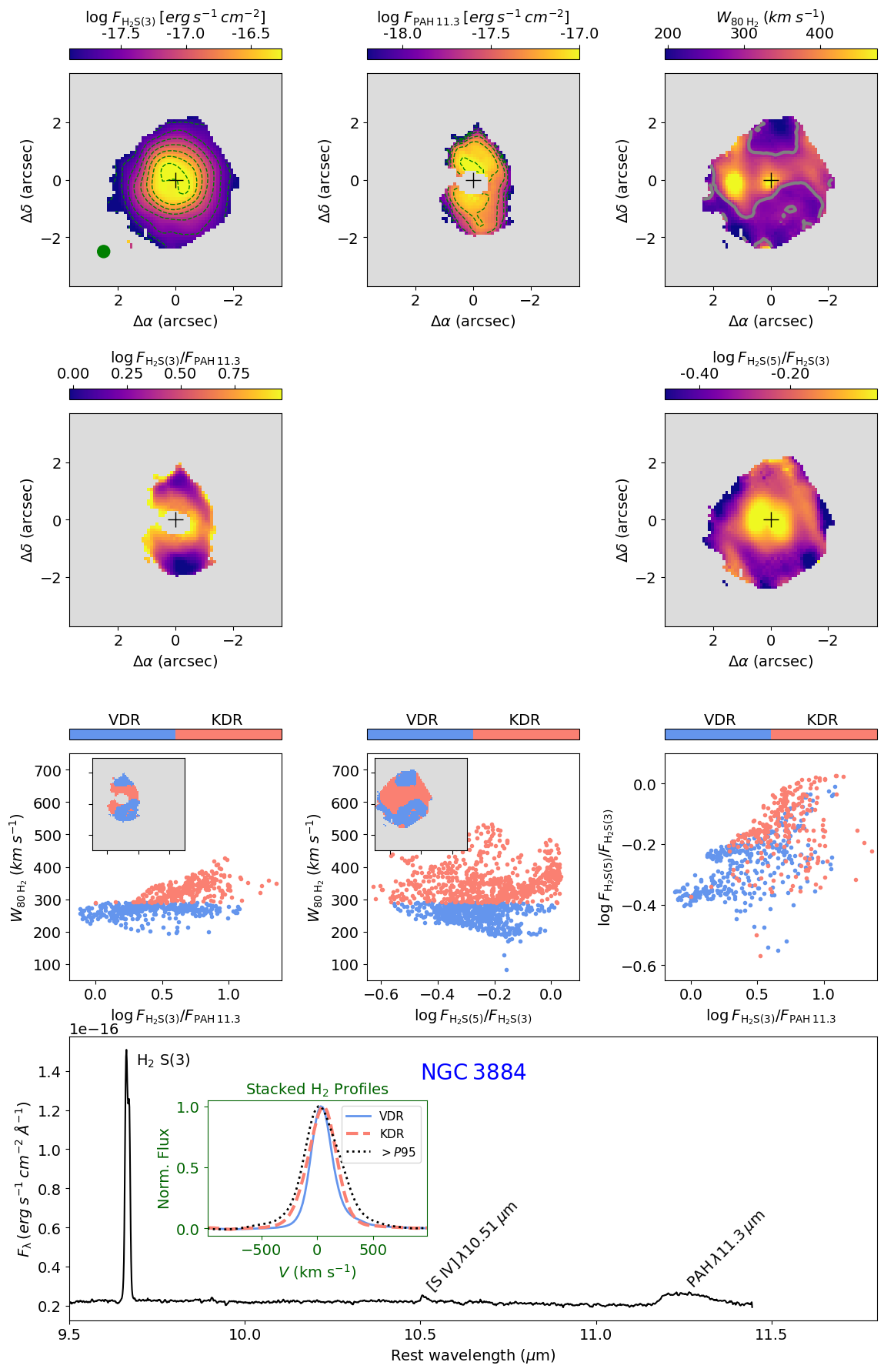}
    \caption{Same as Fig.~\ref{fig:ugc}, but for NGC\,3884. The [Fe\,{\sc ii}]$_{\rm 5.34\: \mu m}$/PAH$_{\rm 11.3\mu m}$ line ratio map is not shown for this galaxy, because the observed spectral range does not include the [Fe\,{\sc ii}]$_{\rm 5.34\: \mu m}$ emission line. Radio contours are not shown as there are no high resolution radio data available for this object.}
    \label{fig:ngc}
\end{figure*}

The results for NGC\:3884 are shown in Fig.~\ref{fig:ngc}. As this galaxy was observed in the MIRI MRS sub-bands Medium and Long, instead of Short and Long, as for the other two galaxies, its integrated spectrum does not include the region beyond 11.5\:$\mu$m. But, the covered spectral range is still enough to include the  PAH$_{\rm 11.3\mu m}$ feature at the galaxy's redshift. The H$_2$ emission in NGC\:3884 is observed over the whole FoV, with the peak of emission seen at the nuclear position. On the other hand, the PAH$_{\rm 11.3\mu m}$ emission is detected only in a partial ring with inner and outer radii of $\sim$0.3 arcsec and 1.8 arcsec, respectively. This is the galaxy with the highest H$_2$ S(3)/PAH$_{\rm 11.3\mu m}$ among the three, with values of up to $\sim$10, much larger than the median values observed in AGN hosts observed by the Spitzer telescope (Fig.~\ref{fig:spitzer}). The observed spectral range also does not include the [Fe\,{\sc ii}]$_{\rm 5.34\: \mu m}$ emission line. The H$_2$ S(5)/H$_2$ S(3) presents the highest values in the inner 0.5 arcsec from the nucleus in a structure elongated along the east west direction. The highest  $W_{\rm 80\:H_2}$ values ($\gtrsim$400~km\:s$^{-1}$) are seen at the nucleus, and along the east-west direction. Regions to the north and south of the nucleus present low $W_{\rm 80\:H_2}$ values ($\lesssim$200~km\:s$^{-1}$).

\section{Discussion}

In this work, we investigate the origin of the excess H$_2$ emission in three galaxies: 
3C\:293, which hosts a radio-loud AGN, and NGC\:3884 and CGCG\:012-070, both hosting radio-quiet AGNs \citep{Becker1995,liu02,Lofthouse18,Lin18,Kukreti22}.

Strong mid-IR H$_2$ emission has been previously reported in radio-loud AGN hosts. For instance,  Spitzer observations of the FR II radio galaxy 3C\:326 reveal exceptionally strong H$_2$ shock-heated emission from its northern component, with the total integrated luminosity of the H$_2$ pure rotational transitions corresponding to 17\:\% of the galaxy's 8-70\:$\mu$m luminosity \citep{Ogle07}. The importance of the radio jet in powering the outflow in 3C\:326N and line emission through interactions between different phases of the ISM gas is reinforced by \cite{Nesvadba10}, using CO (1-0) interferometric and Spitzer observations. They conclude that AGN feedback potentially lowers the star formation efficiency in a way that is consistent with observations of other H$_2$-luminous radio galaxies, with small ratios of CO and PAH surface brightness. \cite{Ogle10}, using a sample of 55 radio galaxies at redshift $z<0.22$ (including 3C 293) observed with the Spitzer satellite, found that 31\:\% of them exhibit pure-rotational H$_2$ emission lines, with molecular masses reaching up to $2\times10^{10}\:{\rm M_\odot}$ and $L_{\rm H_2}/L_{\rm PAH\:7.7\:\mu m}$ ratios approximately 300 times higher than those in normal star-forming galaxies. The authors suggest that the H$_2$ emission in these radio-selected molecular hydrogen emission galaxies (MOHEGs) is caused by heating from radio jets. This conclusion is reinforced by deep Spitzer observations of radio galaxies with neutral and ionized gas outflows (including 3C 293) show that the radio jet, which drives the outflow, is also responsible for the shock excitation of the warm H$_2$ gas \citep{Guillard12}. Modeling the spectral energy distributions from ultraviolet to far-infrared of radio galaxies (including 3C 293) -- selected for the presence of shocked, warm molecular hydrogen emission -- indicates that star formation is suppressed due to shocks driven by the radio jets, which also powers the luminous warm H$_2$ line emission \citep{Lanz16}. Chandra observations of 3C 293 reveal X-ray emission from the jets both within the host galaxy and along the 100 kpc radio jets. The thermal X-ray and warm H$_2$ luminosities are comparable, indicating similar masses of X-ray hot gas and warm molecular gas. This suggests that both components originate from a multiphase, shocked ISM \citep{lanz15}. Multiwavelength imaging and spectroscopy of M\:58 indicate that much of the molecular gas is shocked in situ within the radio jet cocoon, supporting a scenario where jet-driven shocks significantly reduce star formation in the central region of galaxies \citep{Ogle24}. 

The studies above reinforce the importance of shocks driven by radio jets in the production of extremely strong H$_2$ rotational lines in radio galaxies, which can now be studied in more details with JWST.  MIRI MRS observations of IC 5063, a galaxy where the radio jet intercepts the gaseous disk, reveal that the warm H$_2$ is most excited in regions that are co-spatial with the radio structures \citep{Dasyra24}. MIRI MRS observations also revealed jet-driven outflows in warm molecular gas in the inner central regions of 3C\:326\:N \citep{Leftley24} and 3C\:293 \citep{henrique24}, with evidence of shock-heated H$_2$ gas, associated to radio emission.

Shocks also appear to play a significant role in the production of H$_2$ emission in nearby Seyfert and LINER host galaxies. For instance, \citet{Ogle14} found that the rotational H$_2$ emission from the Seyfert host NGC 4258 originates from its inner anomalous arms, which is a signature of jet interaction with the galaxy disk. Additionally, they report a large $L_{\rm H_2}/L_{\rm PAH\:7.7\:\mu m}$ of 0.37, typical of MOHEGs, indicating shocked molecular gas emission. \citet{Garcia-bernete24} use MIRI/MRS observations of the inner $\sim$kpc of three nearby AGN hosts from the  Galactic Activity, Torus, and Outflow Survey (GATOS) map the PAH and H$_2$ MIR flux distributions. They find spatially resolved differences of the H$_2$/PAH ratio for AGN and star-forming dominated regions, with the highest values observed in AGN dominated regions and the lowest ones in the SF dominated regions. Additionally, the  AGN radiation and outflows may influence the PAH population at nuclear and kpc scales, particularly regarding the ionization state of the PAH grains, although their molecular sizes remain relatively similar \citep{Garcia-bernete24}.

The importance of shocks in driving H$_2$ MIR emission in nearby galaxies is further supported by studies of gas kinematics through optical spectroscopy and radio observations. In addition to exhibiting higher H$_2$/PAH ratios, AGN host galaxies also show elevated H$_2$ excitation temperatures compared to normal galaxies \citep[e.g.][]{Petric18,lambrides19}.  \citet{rogemar20_spitzer} investigated the optical properties of a sample composed of 309 nearby galaxies from a parent sample of  2015 objects observed with the Spitzer Telescope, from \citet{lambrides19}. These authors found that the kinematics of [O\:{\sc i}]$\lambda$6300 emission in AGN hosts cannot be explained solely by gas motions resulting from the gravitational potential of their host galaxies, requiring an outflow component. They also observed a correlation between the H$_2$\:S(3)/PAH\:11.3\:$\mu$m ratio and the [O\:{\sc i}] line width, as well as with the [O\:{\sc i}]/H$\alpha$ ratio, a well-known tracer of shocks in neutral gas. This provides an indirect evidence that the H$_2$ emission excess, observed in AGN hosts, is associated to shocks in the same clouds that produce the [O\:{\sc i}] emission \citep{rogemar20_spitzer}. Spatially resolved ALMA observations of cold molecular gas in individual sources reveal that highly disturbed cold gas is closely associated with strong H$_2$ emission and is co-spatial with radio jets \citep{Nesvadba21,Ogle24}. This indicates that the H$_2$ emission is generated by shocks resulting from the energy injected into the ISM by radio jets and outflows. Evidence of shock-heated gas emission in galaxies with H$_2$ emission excess is also observed in hot molecular, neutral and ionized gas through optical and near-infrared integral field spectroscopy \citep[e.g.][]{colina15,rogemar_N1275,rogemar23_ugc,rogemar_n3884}

Except for the results obtained with JWST, most of the findings discussed above are based on integrated observations or imaging of H$_2$ transitions, with limited kinematic information, or rely on indirect tracers such as cold molecular gas observed in the radio or neutral and ionized gas in the optical. Our results allow for a detailed discussion of the origin of H$_2$ emission in the three observed galaxies, mapping both the flux distribution and kinematics in a spatially resolved manner.
For the three galaxies studied here, the maps of the H$_2$ S(3)/PAH$_{\rm 11.3\mu m}$ and [Fe\,{\sc ii}]$_{\rm 5.34\: \mu m}$/PAH$_{\rm 11.3\mu m}$ line ratios exhibit similar patterns among them, for each object. The [Fe\:{\sc ii}] emission is a well-known indicator of shocks. Shocks destroy dust grains, releasing iron (Fe) and increasing its abundance, thereby enhancing its emission \citep[e.g.][]{Oliva01,sb2009_n4151,rogerio13,rogemar_mrk1066_exc,hill14,colina15,rogemar_21_exc}. In addition, these galaxies were selected as having high values of [O\:{\sc i}]$\lambda$6300/H$\alpha$ ratios, with  log\:[O\:{\sc i}]$\lambda$6300/H$\alpha\approx-0.69$ for 3C\:293, log\:[O\:{\sc i}]$\lambda$6300/H$\alpha\approx-0.51$ for CGCG\:012-070, and log\:[O\:{\sc i}]$\lambda$6300/H$\alpha\approx-0.38$ for NGC\:3884 \citep{rogemar20_spitzer}. The [O\:{\sc i}]$\lambda$6300 is also a well-known shock tracer in neutral gas \citep{monreal-ibero06,monreal-ibero10,rich11,rich15}, with shocks being the dominant excitation mechanism for galaxies with log\:[O\:{\sc i}]$\lambda$6300/H$\alpha\gtrsim-1.5$ and [O\:{\sc i}]$\lambda$6300 velocity dispersion ($\sigma_{\rm OI}$) of at least 150 km\:s$^{-1}$ \citep[e.g.][]{allen08,ho14}. The three galaxies have $\sigma_{\rm OI}>220$~km\:s$^{-1}$, as measured from their SDSS spectra \citep{rogemar20_spitzer}. Therefore, these results indicate that, at least partially, the observed H$_2$ line emission in these galaxies is likely due to H$_2$ molecule excitation by shocks.

The plots of $W_{\rm 80 H_2}$ vs. log\:H$_2$ S(3)/PAH$_{\rm 11.3\mu m}$ shown in Figs.~\ref{fig:ugc}, \ref{fig:cgcg} and \ref{fig:ngc} clearly present two trends for all galaxies: (i) one group of points where the H$_2$ S(3)/PAH$_{\rm 11.3\mu m}$ ratio increases with $W_{\rm 80 H_2}$, and (ii) another group of points where the ratio remains practically constant. A correlation between the $W_{\rm 80 H_2}$ with H$_2$ S(3)/PAH$_{\rm 11.3\mu m}$ may indicate a kinematic evidence that shocks are producing the H$_2$ emission excess observed in these galaxies.  To roughly distinguish between the two groups of points and spatially locate them within the galaxies,  we follow \citet{rogemar23_extended_kin} and identify the kinematically disturbed regions (KDRs) and the  virially dominated region (VDRs), using the  $W_{\rm 80}$  measurements. When gas motion is primarily driven by the gravitational potential, the velocity dispersion is expected to decline with increasing distance from the nucleus. Consequently, the velocity dispersion observed at the nucleus can be interpreted as the upper limit of what is attributable to the gravitational potential.  We establish a threshold to define the KDR, based on the fitting of the H$_2$ S(3) emission-line nuclear profile. For each galaxy, we model the H$_2$ S(3) emission line in the nuclear spaxel using two Gaussian components: a narrow component attributed to gas in  the plane of the disc, and a broad component associated with outflows. The KDR correspond to locations with $W_{\rm 80}$ values larger than that of the nuclear narrow component, while other locations  corresponds to the VDR, defined by spaxels with  $W_{\rm 80}$  values smaller than the nuclear one.  The $W_{\rm 80}$  limits are 370\:km\:s$^{-1}$ for 3C\:293,  290\:km\:s$^{-1}$ for NGC\:3884 and 385\:km\:s$^{-1}$ for CGCG 012-070.  The uncertainties in the values of the nuclear narrow component $W_{\rm 80}$ are less than 20 km\:s$^{-1}$ for all three galaxies, resulting in an effect of less than $\sim$5 \% on the fraction of spaxels that define the KDR in all objects. As discussed in \citet{rogemar23_extended_kin}, for nearby AGN hosts, the different assumptions used to select spaxels and calculate the mass of the outflow gas lead to discrepancies of up to 1 dex in the ionized gas outflow rates between different methods, with approximately 0.3 dex of this discrepancy due to the limits $W_{\rm 80}$ used to define the spaxels dominated by outflows.

The $W_{\rm 80}$ cuts are below the escape velocity for all three galaxies. Using SDSS-based stellar masses \citep{maraston09} and isophotal radii \citep{sdss_Dr6}, and assuming that the stellar mass corresponds to 15\% of the total mass of the galaxies, we obtain rough escape velocities of 460 km\:s$^{-1}$ for 3C 293, 540 km\:s$^{-1}$ for NGC\:3884, and 445 km\:s$^{-1}$ for CGCG\:012-070, estimated using a simple Keplerian approximation. For all three galaxies, the $W_{\rm 80}$ limits used to separate the KDR from the VDR are lower than the escape velocities. In 3C\:293, 26\% of the KDR spaxels have velocities above the escape velocity, which accounts for $\sim$51\% of the H$_2$ S(3) emission line flux in the KDR. None of the spaxels in NGC\:3884 exceed the escape velocity, while 62\% of the KDR spaxels in CGCG\:012-070 have velocities above the escape velocity, accounting for 33\% of the H$_2$ S(3) KDR emission.

For 3C\:293, there is a clear trend of $W_{\rm 80 H_2}$ increasing with log\:H$2$ S(3)/PAH${\rm 11.3\mu m}$ in the KDR (Fig.~\ref{fig:ugc}). A similar behavior is observed in the other two galaxies, although it is less pronounced (Figs.~\ref{fig:cgcg} and ~\ref{fig:ngc}).  The increase in the H$_2$/PAH ratio with the H$_2$ line width may indicate that shocks play a significant role in the excitation of the H$_2$ molecule, as larger line widths trace more turbulent gas. This is consistent with results obtained for AGN hosts using Spitzer observations and indirect indicators of gas turbulence derived from optical and radio emission lines, commonly attributed to shocks associated with radio jets \citep[e.g.][]{Nesvadba10,Nesvadba21,Ogle10,Ogle24,rogemar20_spitzer} and by recent MIRI MRS observations of nearby AGN hosts \citep{Leftley24,Garcia-bernete24,Dasyra24}. 

The H$_2$ S(5)/H$_2$ S(3) emission line ratio is a tracer of the gas temperature, representing a fraction of hotter-to-colder molecular gas. In all galaxies, the highest values of this ratio is observed is seen co-spatial with the highest H$2$ S(3)/PAH${\rm 11.3\mu m}$ regions. For  3C\:293 there is a clear correlation between the H$_2$ S(5)/H$_2$ S(3) and H$_2$ S(3) line width and the log\:H$2$ S(5)/H$2$ S(5) vs. log\:H$2$ S(3)/PAH${\rm 11.3\mu m}$ plot shows a similar bimodal behaviour as seen in   $W_{\rm 80 H_2}$ vs. log\:H$_2$ S(3)/PAH$_{\rm 11.3\mu m}$ plot roughly separating the VDR and KDR. In addition, the highest values of all line ratios and $W_{\rm 80 H_2}$ are co-spatial with the radio core \citep{Kukreti22}. This indicates that the outflows in 3C\:293 are driven by the radio jet -- as previously reported \citep{mahony16,rogemar23_ugc,henrique24} and the H$_2$ emission from the KDR is produced by shock excitation. CGCG 012-070 and NGC\:3884 also show the highest H$_2$ S(5)/H$_2$ and H$_2$ S(3)/PAH${\rm 11.3\mu m}$ line ratios in co-spatial regions, with a trend suggesting that the increases in both ratios are closely linked. These two galaxies do not show a clear trend of higher H$_2$ S(5)/H$_2$ S(3) ratios (indicating hotter gas) being associated with line widths (more turbulent gas), and they also do not have radio images with similar resolution to our data, despite being detected by the VLA FIRST survey \citep{Becker1995,Lofthouse18,Lin18}. This suggests that the contribution of shocks to the excitation of H$_2$ mid-infrared lines is less significant compared to 3C\:293, but still not negligible for the two galaxies, in  particular for the KDR.


The insets in the bottom panels of Figs~\ref{fig:ugc}, \ref{fig:cgcg} and \ref{fig:ngc} show stacked H$_2$ S(3) emission line profiles for the KDR (in salmon) and in the VDR (blue) for the three galaxies. These profiles were created by summing all spaxels in each region, using the centroid velocity of the line as the reference, and then normalizing them by the peak of the profile. The line profiles in the KDRs are more asymmetric than those in the VDRs. In the case of CGCG 012-070, the line profiles from both regions show similar widths and exhibit a blueshifted asymmetric wing. Although the distribution in the $W_{\rm 80 H_2}$ versus H$2$ S(3)/PAH${\rm 11.3\mu m}$ plot is bimodal for this galaxy (Fig.~\ref{fig:cgcg}), correlations between these parameters are observed in both groups of points. This suggests that shocks contribute additionally in the VDR of CGCG 012-070.

\begin{figure*}
    \centering
    \includegraphics[width=0.79\textwidth]{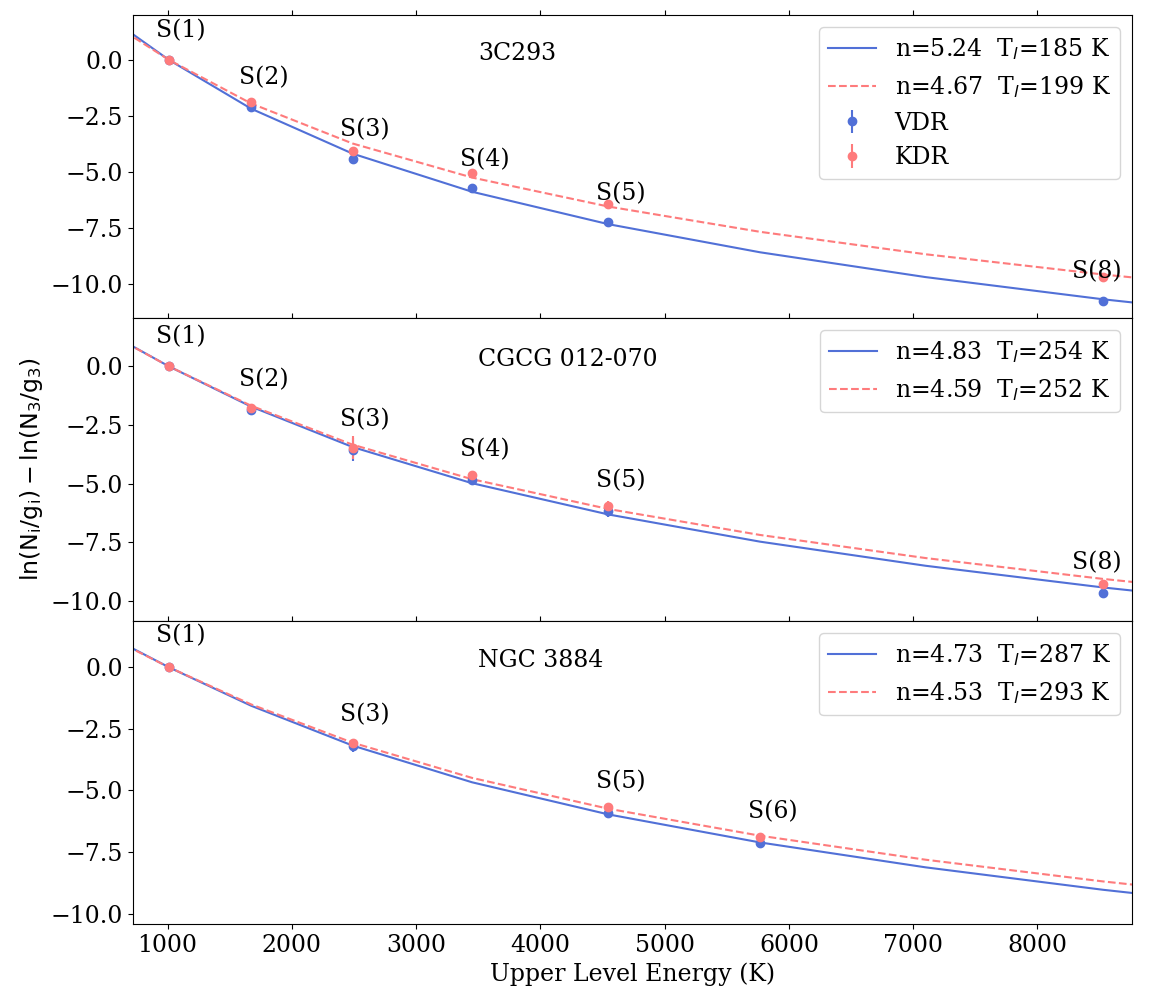}
    \caption{Excitation diagrams for the VDR and KDR in 3C\:293 (top), CGCG\:012-070 (middle) and NGC\:3884 (bottom). The points are obtained from the integrated H$_2$ fluxes, while the lines correspond to the best-fitted power-law temperature model.}
    \label{fig:temperature}
\end{figure*}

We can further investigate the role of shocks in the excitation of the H$_2$ gas, using the observed luminosities of the H$_2$ emission lines and adopting a power-law temperature distribution \citep{zakamska10,Pereira-Santaella14,togi16}. We follow the procedure described in \citet{henrique24}, evaluating the column density of the $J$ state ($N_J$) by 
    \begin{equation} \label{eq:NJ}
        N_{J} = \int_{T_{u}}^{T_{l}} f(T)m T^{-n}d T,
    \end{equation}
where $m$ is a normalization constant $f(T)$ is the Boltzmann factor, $T_{u}$  and $T_{l}$ are the upper and lower temperature limits and $n$ is the power-law index. Figure~\ref{fig:temperature} shows the excitation for the three galaxies in our sample, using all the observed H$_2$ transitions detected in their spectra. The plots display on the {\it y}-axis the population number of the $i$ state divided by the state degeneracy and normalized by the $i=3$ state, while the {\it x}-axis represents the upper level energy. We compute the integrated fluxes for each H$_2$ emission line for the KDR and  VDR, considering spaxels where all emission lines are detected.  We corrected the observed fluxes for extinction using the G23 extinction law  \citep{Gordon23,Gordon21,Gordon09,Gordon24,Decleir22,Fitzpatrick19} and mean visual extinctions ($A_{\rm V}$) of 2.6, 1.4, and 0.8 mag, estimated from the Balmer decrement using H$\alpha$ and H$\beta$ flux measurements from \citet{Thomas13} for 3C\:293, CGCG\:012-070, and NGC\:3884, respectively.
 The best fit models are shown as salmon and blue lines for the  KDR  and  VDR, respectively. To obtain the best model, we kept $T_u$\,=\,5000\:K fixed \citep{Yuan11} and varied the other two parameters, $T_l$ and $n$. The best-fit parameters are shown in the top-right corner of the plots, for each galaxy. 

If molecular gas is heated by shocks, a flat power-law index is expected, resulting in higher temperatures that favor a larger fraction of gas being in hotter phases compared to gas in photodissociation regions \citep{togi16}. For instance, \citet{togi16} reported power-law index values in the range 3.79--6.39, and a mean value of $\langle n \rangle=4.84\pm0.61$, for a sample of 43 galaxies from the Spitzer Infrared Nearby
Galaxies Survey \citep[SINGS, ][]{Kennicutt03}, which include LINERs, Seyferts, Dwarfs, and Star-forming galaxies. Considering only the Star-formation dominated galaxies (18) in their sample, the mean value is $\langle n \rangle=5.16\pm0.36$, while considering only AGN (Seyferts and LINERs) hosts (20) the mean index is $\langle n \rangle=4.46\pm0.44$. \citet{Appleton17} present a two-dimensional map of the power-law index for the Stephan’s Quintet, based on Spitzer images. They found values in the range $\sim$4.2--5.0, with the lowest values observed in shock dominated regions. The same range of $n$ values is reported by \citet{Appleton23} using JWST MIRI observations. ULIRGs typically exhibit power-law index values within the range 2.5--5.0, with lower values associated to shock dominated regions \citep[e.g.][]{zakamska10}. As shown in Figure~\ref{fig:temperature}, the KDR in the three galaxies exhibits lower $n$ values compared to the VDR, with the largest difference seen for 3C\:293. This confirms that shocks play a major role in the observed emission from the kinematically disturbed region.

We can speculate on the origin of the shocks. The three galaxies studied here include different representatives of AGNs regarding the radio emission and optical classification. NGC\,3884 hosts a type 1 LINER, CGCG\,012-070 hosts a Sy\,2 nucleus, and 3C\,293 hosts a LINER \citep{veron06}.   3C\,293 is a radio loud source, whereas NGC\,3884 and CGCG\,012-070 are radio quiet, but presenting faint radio emission \citep{liu02,Becker1995,Lofthouse18,Lin18}. In addition these galaxies present outflows in ionized gas, which seem to be mechanically driven by the interaction of the radio jet with the ambient gas in the 3C\,293 and likely radiatively driven in the other two, as they show only faint radio emission \citep{rogemar23_ugc,rogemar_n3884}. The shock-dominated regions are seen mostly in the central region for all galaxies. In addition, the shock-dominated region in 3C\:293 is seen co-spatial with the radio core, indicating that the origin of the shocks is the interaction of the radio jet with the ambient gas, in agreement with previous results using Spitzer observations and indirect tracers of the warm H$_2$ emission \citep{Ogle10,Guillard12,lanz15}. The jet-cloud interaction is also responsible for driving the warm molecular and ionized gas outflows in this object \citep{mahony16,rogemar23_ugc,henrique24}.  Since NGC\:3884 and CGCG\:012-070 exhibit only weak radio emission, the origin of the shocks in these galaxies is likely due to radiatively driven outflows. However, because of the lack of high-sensitivity radio observations for these galaxies, low-power jets cannot be ruled out as the potential source of their shock-heated H$_2$ emission.  Indeed, interactions between low-power radio jets and the interstellar medium are the source of the outflows observed in ionized gas in nearby AGN hosts \citep{nandi2023,girdhar2024}. This interpretation aligns with predictions from simulations indicating that molecules can form within outflows, with their emission attributed to shock heating \citep{richings18a,richings18b}.

\section{Conclusions}
We used JWST MIRI MRS observations of the inner $\sim$2 kpc of the nearby AGN hosts 3C\,293, CGCG\,012-070 and NGC\,3884 to investigate the origin of the H$_2$ emission excess observed in these galaxies. Our main conclusions are:
\begin{itemize}
    \item The median values of log\:H$_2$ S(3)/PAH$_{\rm 11.3\mu m}$ observed for 3C\:293, CGCG\:012-070, and NGC\:3884 are 0.15, 0.35, and 3.32, respectively. These values significantly exceed those measured for non-active galaxies using Spitzer data, which have a median value of log\:H$_2$ S(3)/PAH$_{\rm 11.3\mu m} = -1.74$, as well as the median values for AGN hosts, which are typically around log\:H$_2$ S(3)/PAH$_{\rm 11.3\mu m} = -1.26$. This places these galaxies among the most extreme cases of H$_2$ emission excess.
    \item  Plots of $W_{\rm 80 H_2}$ vs. log\:H$_2$ S(3)/PAH$_{\rm 11.3\mu m}$ clearly exhibit two trends in all galaxies: (i) one cluster of points where the H$_2$ S(3)/PAH$_{\rm 11.3\mu m}$ ratio and $W_{\rm 80 H_2}$ present a strong correlation, and (ii) another cluster with the ratio remaining nearly constant. 
    \item Using the $W_{\rm 80 H_2}$ values, we define the kinematically disturbed region (KDR) and the virially dominated region (VDR), and find that shock heating of the gas is the primary excitation mechanism for H$_2$ emission in the KDR. This is supported by larger fractions of hotter-to-colder molecular gas (traced by the H$_2$ S(5)/H$_2$ S(3) ratio), a correlation between the H$_2$ and [Fe\:{\sc ii}] and flatter power-law temperature distributions in the KDR, as compared to the VDR.

    \item The origin of the shocks in 3C 293 is the interaction between the radio jet and the interstellar medium, as indicated by the spatial correlation between the radio emission and the highest line ratio values and warm molecular jet-driven outflows.  The other two galaxies exhibit modest radio emission; however, the interaction of low-power jets with the interstellar medium cannot be ruled out as a potential source of the shocks. Higher sensitivity radio observations are necessary to further investigate the role of the radio jet in the origin of the excess H$_2$ emission in these galaxies. 
\end{itemize}

Our work demonstrates the capability of the JWST MIRI MRS in investigating the origin of H$_2$ emission excess in nearby AGN hosts, as revealed by Spitzer observations. While Spitzer's limited spatial and spectral resolutions constrained our ability to disentangle shock and AGN heating processes, JWST now provides the detailed spatial and spectral information necessary to understand the formation and origins of molecular emission, as well as its relationship with outflows.

\section*{acknowledgments}
We would like to thank an anonymous referee for their contributions, which greatly improved the presentation of this manuscript. RAR acknowledges the support from Conselho Nacional de Desenvolvimento Cient\'ifico e Tecnol\'ogico (CNPq; Proj. 303450/2022-3, 403398/2023-1, \& 441722/2023-7), Funda\c c\~ao de Amparo \`a pesquisa do Estado do Rio Grande do Sul (FAPERGS; Proj. 21/2551-0002018-0), and Coordena\c c\~ao de Aperfei\c coamento de Pessoal de N\'ivel Superior (CAPES;  Proj. 88887.894973/2023-00).  HCPS and GLSO thank the financial support from CAPES (Finance Code 001) and CNPq.   N.L.Z. is supported in part by NASA through STScI grant JWST-ERS-01928.
 RR acknowledges support from CNPq (Proj. 311223/2020-6,  304927/2017-1, 400352/2016-8, and  404238/2021-1), FAPERGS (Proj. 19/1750-2 and 24/2551-0001282-6) and (CAPES, Proj. Proj. 88887.894973/2023-00). MB thanks the financial support from the IAU-Gruber foundation fellowship.

 \section*{Data Availability}
The data used in this work are part of the JWST cycle 1 project (ID 1928). The complete dataset can be accessed at the Mikulski Archive for Space Telescopes (MAST) platform at the Space Telescope Science Institute, through \dataset[10.17909/tazj-hp44]{https://doi.org/10.17909/tazj-hp44}.

\facilities{JWST (MIRI MRS), MAST, NED.}

\bibliography{riffel}{}
\bibliographystyle{aasjournal}



\end{document}